\documentclass{article}
\usepackage{epsfig}
\begin{document}

\centerline{\bf \Large A 233 km Tunnel for Lepton and Hadron Colliders}
\bigskip
\bigskip

\centerline{\large D. J. Summers, L. M. Cremaldi, A. Datta, M. Duraisamy, T. Luo, G. T. Lyons}
\medskip
\centerline{\it \footnotesize Dept. of Physics and Astronomy, University of Mississippi-Oxford, University, MS 38677 \ USA}

\vspace{-4mm}

\renewcommand{\abstractname}{}
\begin{abstract} \footnotesize \noindent {\bf Abstract.} A decade ago, a cost analysis was conducted to bore a 233 km circumference 
Very Large Hadron Collider (VLHC) tunnel passing through Fermilab. 
Here we outline implementations of $e^+e^-$, $p \bar{p}$, and $\mu^+ \mu^-$ 
collider rings in this tunnel using recent technological 
innovations. The 240 and 500 GeV $e^+e^-$ colliders employ  Crab Waist Crossings, ultra low 
emittance damped bunches, short vertical IP focal lengths, 
superconducting RF, and low coercivity, grain oriented silicon 
steel/concrete dipoles. Some details are also provided for a high luminosity 
240 GeV $e^+ e^-$ collider and 1.75 TeV muon accelerator in a Fermilab site 
filler tunnel. The 40 TeV $p \bar{p}$ collider uses the high intensity 
Fermilab $\bar{p}$ source, exploits high cross sections for $p \bar{p}$
production of high mass states, and uses 2 Tesla
ultra low carbon steel/YBCO superconducting magnets run with liquid neon. 
The 35 TeV muon ring ramps the 2 Tesla superconducting magnets at 9 Hz 
every 0.4 seconds, uses 250 GV of superconducting RF to accelerate muons 
from 1.75 to 17.5 TeV in 63 orbits with 71\% survival, and mitigates 
neutrino radiation with  phase shifting, roller coaster motion  in a FODO lattice. \newline
{\scriptsize {\bf PACS:} electron positron collider, proton antiproton collider, muon collider}
\end{abstract}

%%%%%%%%%%%%%%%%%%%%%%%%%%%%%%%%%%%%%%%%%%%%
%% MAINMATTER
%%%%%%%%%%%%%%%%%%%%%%%%%%%%%%%%%%%%%%%%%%%%

\bigskip
\bigskip
\bigskip
\centerline{\bf  \large INTRODUCTION}
\medskip

{\footnotesize
\noindent
In 2001, a cost estimate\,\cite{CNA} for boring a 233 km circumference tunnel in 
northern Illinois was made for the proposed Very Large Hadron Collider (VLHC\,\cite{Ambrosio}).
Level 12 foot and 16 foot diameter tunnels were estimated to cost  \$2.55 billion and \$2.94 billion, respectively.
Included were a shotcrete lined tunnel, caverns, vertical access shafts, and 25\% contingency.
Since then inflation has increased prices, but more 
automation has been added to tunneling in pulling tunnel boring machines 
forward and in placing rock stabilization bolts\,\cite{Lyons}. 
Here we outline how such a tunnel might be used by lepton and   hadron colliders over many years.
We examine circular 240 and 500 GeV $e^+ e^-$ colliders\,\cite{Lyons, Sen}, a 40 TeV $p \, \bar{p}$ collider, and a 35 TeV $\mu^+ \mu^-$ collider\,\cite{Lyons, Neuffer}.
The recent observation of a 126 GeV/c$^2$ boson\,\cite{Boson} provides motivation for  240 GeV $e^+ e^- \!\to Z^{\,0} h^0$\,\cite{Blondel} and 
126 GeV $\mu^+ \mu^- \to h^0$\,\cite{Neuffer2} colliders . 

\bigskip
\bigskip
\centerline{\bf \boldmath  \large 240 and 500 GeV  \boldmath $e^+ e^-$ Ring Colliders}
\medskip

A crab waist crossing\,\cite{Raimondi2006} as developed 
for the next generation of B factories is employed to extend the energy of circular $e^+ e^-$ colliders beyond LEP.  Low emittance 
bunches from precision damping rings for the proposed 
International Linear Collider (ILC\,\cite{Brau}) are used as well the short vertical focal 
length ILC collision region optics. 

A beam crossing angle is introduced to allow short focal length, $\beta_y^*$, collision optics.  
The horizontal emittance of the beam is driven by quantum fluctuations in synchrotron radiation\,\cite{Lee}.
The vertical emittance is lowered until the tune-tune shift limit, $\xi_y$,  is reached.  The crossing angle independent luminosity is given by\,\cite{Lyons, Raimondi2006}:

\begin{equation}
L = 2.167 \times 10^{\,34}  \  {\rm{E} \,(GeV) \ I\,(Amps)} \   \xi_y / \beta^*_y\,\hbox{(cm).}
\end{equation}

Preliminary parameters for three high energy, high luminosity machines are given in Table 1.
One of  the 240 GeV machines fits in a Fermilab site filler ring. A 120\,mm bore Nb$_3$Sn  quadrupole\,\cite{Bossert} may be useful in getting the beam to fit into the final focus.

However, some beam particles are lost due to beamstrahlung tails when the momentum changes so much that the ring and Interaction Point (IP) 
can no longer transport the particles.
Equation 2 shown below is used to  calculate luminosities under these conditions\,\cite{Telnov}. $L$ is luminosity, $h$ is the hourglass factor, $\eta$ is momentum acceptance,
$\xi_y$ is the vertical beam-beam tune shift, $E_0$ is the beam energy, $\varepsilon_y$ is the vertical geometric emittance, $P$ is the synchrotron radiation power for both beams,
$R$ is the ring radius, and $R_{\rm{b}}$ is the bending radius.  To still attain reasonable luminosities we use a large ring, increase the normal ring/IP $\eta =$ 1\% momentum acceptance
to $\eta =$ 3\% gaining a factor of $3^{2/3} = 2.08$, and employ the ILC vertical emittance.  Luminosity scales linearly with ring circumference for fixed synchrotron power.
There is experience with 3\% momentum acceptance rings\,\cite{Machida}, which would need to be designed to minimize synchrotron radiation losses.
A large momentum acceptance  IP design would be new, probably  allocating more real estate to sextupoles and less to quadrupoles, if it can be built without too much
of an increase in focal length.
The nominal normalized ILC vertical emittance is 0.04 mm-mrad.  This yields geometric emittances  of  0.000170 nm and 0.000082 nm
for 120 GeV ($\gamma$ = 235,000) and 250 GeV ($\gamma$ = 489,000) beams, respectively.  
Results are in Table 2.  The luminosity might also be improved by refreshing the beam at up to a few times per second instead of the the roughly 12 minute interval\,\cite{Blondel}
required by radiative Bhabba scattering.  Table 2 shows refresh times that  that add 10\% to power requirements beyond synchrotron radiation. Finally, Reference\,\cite{Telnov} notes that a crab waist crossing would add a further factor of $2^{2/3} = 1.6$ to the luminosities shown in the last line of Table 2.

\begin{equation}
\frac{L}{10^{34\rule{0pt}{7pt}} \, {\rm{cm}}^{-2} \, {\rm{s}}^{-1}} \approx \frac{100 \, h \, \eta^{2/3} \, \xi_y^{1/3}}{(E_0/100 \, \, {\rm{GeV}})^{13/3} \,
(\varepsilon_y/{\rm{nm}})^{1/3} } 
\times
\left(\frac{P}{100\, \mbox{MW}}\right)
\left(\frac{2\pi R}{100\, \, {\rm{km}}}\right) \frac{R_\mathrm{b}}{R}
\end{equation}

The dipoles for this 233 km circumference ring have a magnetic 
field four times lower than used at the CERN LEP machine.  To maintain good 
field quality, particularly at injection, a soft magnetic material is needed.
Grain oriented silicon steel\,\cite{Shirkoohi} is chosen for
the dipoles because its coercivity is 1/5 that  of ultra low carbon steel\,\cite{Laeger}.
Horizontal bands sandwich the top and bottom of C shaped laminations to permit a high
permeability path in the entire flux return circuit. Putting concrete in between laminations
provides space for the four bands. If an even lower coercivity material is absolutely required, hydrogen annealed mu metal (77\% nickel, 16\% iron) might suffice.

%% Table 1 here

\begin{table}[t!]
\centering
{\footnotesize
   \caption{\footnotesize Strawman parameters for three $e^+ e^-$ colliders which exploit the crab waist crossing\,\cite{Raimondi2006}.}
\smallskip   
\renewcommand{\arraystretch}{1.20}
\begin{tabular}{lcccl} \hline
Parameter  Name\rule{0pt}{10pt} (Units) &  &&& Formulae \\ [0.3ex] \hline 
\boldmath $e^+, e^-$\rule{0pt}{10pt} energy  (GeV) & 120, 120   & 120, 120 & 250,  250 & \\
Ring Circumference: C  (km)               & 15 & 233 & 233 & \\
Ring Radius: R (meters)                              & 2400  & 37,\,100   & 37,\,100  & R = C / 2$\pi$ \\
Bending radius: $\rho$   (meters)          &   1900  &   29,\,000 & 29,\,000 &\\  
Relativistic \boldmath $\gamma$      & 235,\,000 & 235,\,000 &  489,\,000 &E /\,m = (120, 250)\,/0.000511\\
Collision frequency: $f_0$ (kHz) & 65.1 &  978 &  52.8 & (Bunches\,/\,beam) c / 2$\pi$R\\
Half crossing angle: $\theta$ (mr) & 34 & 34 & 34 &\\
Bunch length (mm) & 6.67 &  6.67 & 6.67 & \\
$\sigma_x, \sigma_y$ IP beam size ($\mu$m) & 8.5, 0.0244 &  8.5,  0.0244 &  8.5,  0.0115 & $\sigma = \sqrt{\epsilon\,\beta^*\rule{0pt}{7pt}}$\\
IP $\beta_x^*, \beta_y^*$    (cm) & 2, 0.06 &  2, 0.06 & 2, 0.06 & \\
Geometric emittance:  $\epsilon_x$ (nm) & 3.6  &  3.6 &  3.6 &  $\sim \hbox{(Lattice Type)}\, \gamma^{\,2} {\rm{(\ell_{\hbox{\footnotesize{\,half cell}}}/ \rho})}^3$\,\cite{Lee}\\
Geometric emittance:  $\epsilon_y$ (nm) & 0.00099  & 0.00099 &  0.00022 &  \\
Norm.\,emit.:  $\epsilon_x^N, \epsilon_y^N$ (mm-mrad) & 846, 0.235   &  846, 0.235 &  1760, 0.108 & $\epsilon^N = \gamma \, \epsilon$\\
Beam-beam tune shift: $\xi_x$  & 0.0014 &  0.0014 &  0.0007 & $r_e \, N / 4 \pi \epsilon_x^N \approx 2 \,r_e N \beta_x^* / (\pi \gamma \sigma_x^2 \,\theta^2)$\,\cite{Raimondi2006}\\
Beam-beam tune shift: $\xi_y$  & 0.20 &  0.20 & 0.23\,\cite{Sen} & $r_e\,  N / 4 \pi \epsilon_y^N \approx r_e N \beta_y^* / (2 \pi \gamma \sigma_y \sigma_z \,\theta)\rule{0pt}{9pt}$\,\cite{Raimondi2006}\\
No.~of bunches\,/\,beam      &  3 &  700 &  41 &\\
Particles / bunch\,\cite{Sen} &  $4.85 \times 10^{11}$ &  $4.85 \times 10^{11}$ &   $4.85 \times 10^{11}$ & $\delta N_2 = 2 N_2  \sigma_x / (\theta \sigma_z) = 3.63 \times 10^{10}$\,\cite{Raimondi2006}\\      
Dipole field (Tesla) & 0.21  &  0.014 &  0.029 & B = (120, 250)\,/.3$\rho$\,(meters)\\
Current / beam   (Amps)                       &  0.00505  &  0.07 &  0.0041 & $1.6 \!\times\! 10^{-19}$\,(particles/beam) c /\,2$\pi$R\\
E loss / orbit  (GeV)                             &    9.7  &  0.63 &  11.9 &  $8.85 \times 10^{-5} \, E^4({\rm{GeV}})  / \rho({\rm{m}})$\\
Synch rad power (MW/\,beam)          &  49  &  44 &  49 & $8.85 \times 10^{-2} \, E^4({\rm{GeV}}) \, I({\rm{amps}}) / \rho({\rm{m}})$\\
Total synch wall power (MW)             &  198  &  176 &  198 & \\
IP $\beta^{\,\rm{max}}_x$, $\beta^{\,\rm{max}}_y$ (km)                  & 40, 250 & 40, 250 &  40, 250 & \\
IP \ $\sigma^{\,\rm{max}}_x$, $\sigma^{\,\rm{max}}_y$ (mm)                  & 12, 0.5  &  12, 0.5 &  12, 0.23 & $\sigma^{\,\rm{max}} = \sqrt{\epsilon\,\beta^{\,\rm{max}}\rule{0pt}{7pt}}$ \\
IP Sextupole Strength   (1/m)$^{\,2}$ &   0.0007 &  0.0007 & 0.0007 &  $K_2 = [1/(2 \,\theta \, \beta^{\,\rm{max}}_y \beta^*_y)]   \sqrt{\beta^*_x / \beta^{\,\rm{max}}_x} $\,\cite{Raimondi2006} \\
Luminosity (cm$^{-2}$ s$^{-1}$) &  $4.4 \times 10^{\,34}$  &  $6.1 \times 10^{\,35}$ &  $7.6 \times 10^{\,34}$  &
 L = $N_1 \, ({\delta}N_2) f_0 / (4 \pi \sigma_x \,\sigma_y)\rule[-4pt]{0pt}{5pt}$ \\ [0.3ex] \hline
\end{tabular}
}
\end{table}

 \begin{table}[h]
 \centering
{\footnotesize 
   \caption{\footnotesize Strawman parameters for three $e^+ e^-$ colliders including beamstrahlung tail limitations\,\cite{Telnov}. A  beam refresh time of a second or less
   might improve luminosity but is not included in Equation 2 values on the last line.}
\smallskip   
\renewcommand{\arraystretch}{1.20}
\begin{tabular}{lcccl} \hline
Parameter  Name\rule{0pt}{10pt} (Units) &  &&& Formulae \\ [0.3ex] \hline 
\boldmath $ E_0: \, e^+, e^-$\rule{0pt}{10pt} energy  (GeV) & 120, 120   & 120, 120 & 250,  250 & \\
Ring Circumference: C  (km)               & 15 & 233 & 233 & \\
Ring Radius: R (km)                              & 2.4  & 37.1   & 37.1  & R = C / 2$\pi$ \\
Bending radius: $R_{\rm{b}}$   (km)          &   1.9  &   29 & 29 &\\  
Hourglass factor, h                             & 0.8 & 0.8  & 0.8 &\\
Ring and IP momentum acceptance $\eta$ & 0.03 & 0.03 & 0.03 & \\
Relativistic \boldmath $\gamma$      & 235,\,000 & 235,\,000 &  489,\,000 &E /\,m = (120, 250)\,/0.000511\\
Norm.\,emit.:  $\epsilon_y^N$ (mm-mrad) & 0.04   &  0.04 &  0.04 & $\epsilon^N = \gamma \, \epsilon$\\
Geometric emittance:  $\epsilon_y$ (nm) & 0.000170  & 0.000170 &  0.000082 &  \\
Beam-beam tune shift: $\xi_y$  &0.15 &  0.15 & 0.15 & $r_e\,  N / 4 \pi \epsilon_y^N$\\
E loss / orbit  (GeV)                              &    9.7  &  0.63 &  11.9 &  $8.85 \times 10^{-5} \, E^4({\rm{GeV}})  / R_{\rm{b}}({\rm{m}})$\\
Beam refresh time (seconds)             &   0.006  &  1.5    &  0.16   &   $10 \times (E_0 \, / \, E_{\rm{loss/orbit}}) (C/300,000)$       \\
Synch rad power, both beams (MW)          &  100  &  100 &  100 & $8.85 \times 10^{-2} \, E^4({\rm{GeV}}) \, I({\rm{amps}}) /  R_{\rm{b}}({\rm{m}})$\\
Luminosity (cm$^{-2}$ s$^{-1}$) &  $4.0 \times 10^{\,34}$  &  $6.3 \times 10^{\,35}$ &  $3.3 \times 10^{\,34}$  &
 Equation 2\,\cite{Telnov} \\ [0.3ex] \hline
\end{tabular}
}
\end{table}

\vspace{-3mm}
\begin{figure}[b!]
   \centering
   \includegraphics*[width=80mm]{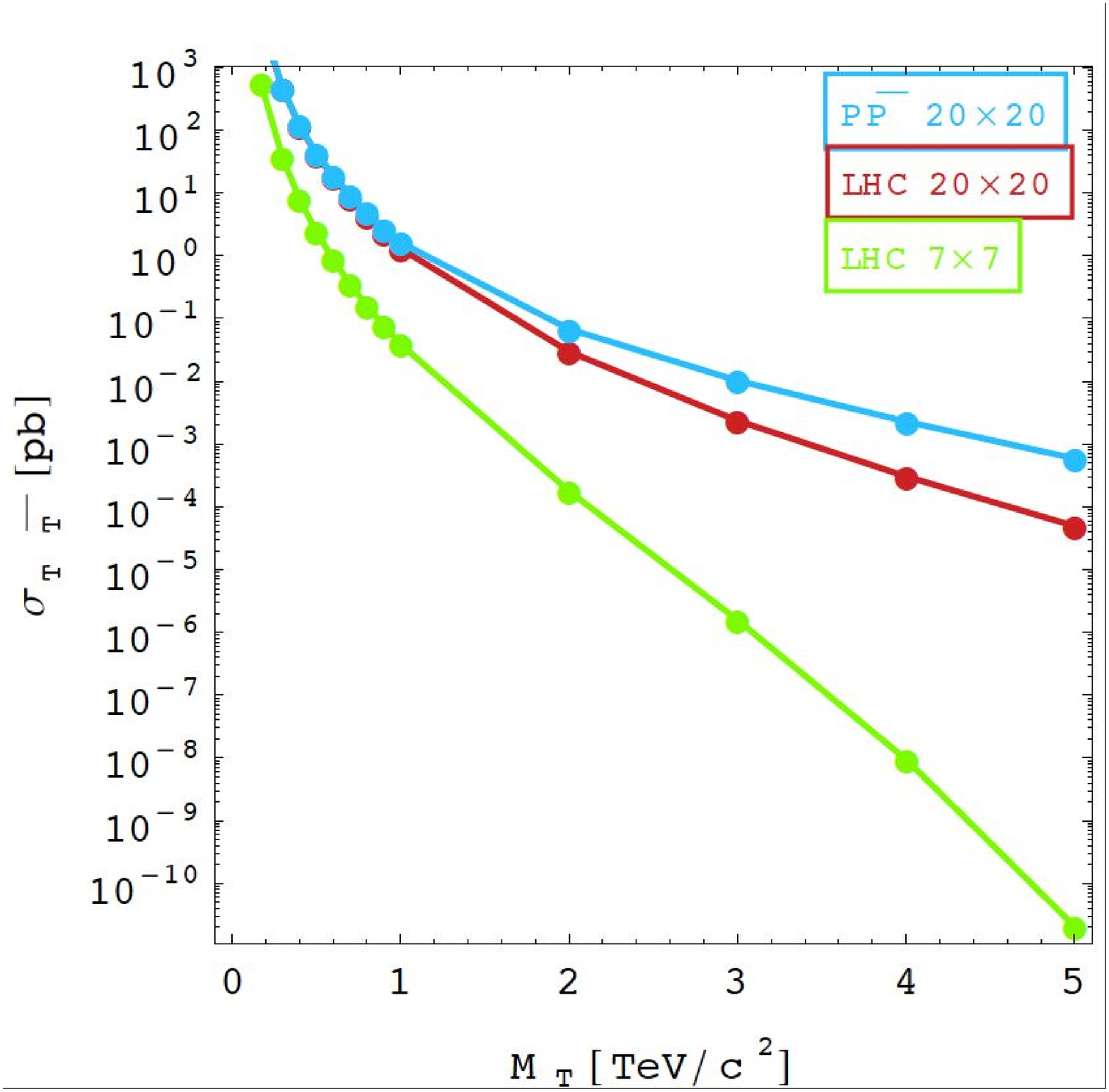}
\vspace{-3mm}   
   \caption{\footnotesize $p\,\bar{p}$ and $p\,p$ cross sections are generated\,\cite{Stelzer} for a particle similar to the top quark as a function of mass.}
\end{figure}

\bigskip
\bigskip
\bigskip
\centerline{\bf \boldmath  \large Energy Frontier 40 TeV  $p \,\bar{p}$ Collider}
\medskip

A 40 TeV $p \, \bar{p}$ collider fits in the 233 km tunnel with 2 Tesla H-frame dipoles.
Ultra low carbon steel\,\cite{Laeger} is used for the dipoles.  The low coercivity/\,hysteresis loss of this steel permits reuse of these magnets  for
a muon collider. The magnet coils consist of 52 turns of 4mm wide YBCO superconducting ribbon.
Each ribbon carries 500 amps for a total of 26,\,000 ampere\,/\,turns.  The coils are cooled with liquid neon at 25K\,\cite{Lyons}.

The Tevatron luminosity\,\cite{Holmes} of  $4 \times 10^{\,32}\, \rm{cm}^{-2} \, s^{-1}$ is scaled to yield:

\begin{equation}
L = (20/37) \,(4 \times 10^{\,32}) = 2.16\times 10^{\,32}\, {\rm{cm}^{-2} \, s^{-1}.}
\end{equation}

The factor of 20 increase comes from the energy increase and the factor of 37 decrease comes from lowering the collision frequency due to  the larger ring.
As shown in Fig.~1, the $p \, \bar{p}$ cross section for many high mass states  is an order of magnitude larger than the $p\,p$ cross section. 
Thus, for high mass objects near threshold, this collider, with the Tevatron $\bar{p}$  source,
%has double the event rate of
has 2x more events and 5x less background than
the Superconducting Super Collider (SSC) $p\,p$ design with a luminosity of
$10^{\,33}\, \rm{cm}^{-2} \, s^{-1}$.  The cross section for $p \, \bar{p}$ collisions does rise from 80 to 120 mb as $\sqrt{s}$ goes from 2 to 40 TeV.
But this increased $\bar{p}$  burn
rate might be ameliorated by adding a second, parallel $\bar{p}$ accumulator ring.
The current
limitation on the Tevatron $\bar{p}$ source is the accumulator ring with a $\bar{p}$ stacking rate of
$26 \times 10^{10}$  $\bar{p}$\,/\,hour\,\cite{Pasquinelli}. The debuncher ring can supply $40 \times 10^{10}$  $\bar{p}$\,/\,hour.

\bigskip
\bigskip
\centerline{\bf \boldmath \large Energy Frontier 35 TeV  $\mu^+ \mu^-$ Collider}
\medskip

First we  calculate the  neutrino radiation ($\,\mu^- \!\to e^- \, {\overline{\nu}}_e \, \nu_{\mu}$ \, and \,
$\mu^+ \to e^+ \, \nu_e \, {\overline{\nu}}_{\mu}$)
for a ring with 17.5 TeV muons\,\cite{King}.
A 17.5 TeV muon lifetime is 0.364\,s and $\gamma$ = 165,000.

\begin{equation}
\tau = \gamma \, \tau_{\mu^{\pm}}  =  {17.5 \, TeV \over 105.7 \, MeV} \  \  2.2 \times 10^{-6}\,\hbox{s} =   0.364 \,\hbox{s}
\end{equation}

\begin{equation}
{D_{exit}^{ave}[{\rm{Sievert}}]} = {2.9 \!\times\! 10^{-24}} \times {{{N_\mu} \  {({E_\mu} [\rm{TeV}])^3}} \over D[\rm{m}]} =      
{2.9 \!\times\! 10^{-24}} \times {{{(1.1 \!\times\! 10^{\,20})} \ {({17.5}\ \rm{TeV)^3}}} \over 300\ \rm{m}} = {0.0057\ {\rm{Sv/ yr}}}  
\end{equation}

The ring is  300\,m underground.
Two  bunches of $2\! \times\! 10^{12}$ muons are produced every 0.364 seconds giving  $1.1 \times 10^{\,20}$ muons per $10^7$ second
accelerator year.  The radiation dose, which  equals the yearly dose from background,  is too high, 0.0057 Sieverts\,/year or 570 mrem\,/year.
A neutrino from the three body decay of a 17.5 TeV/c muon has a 20 MeV/c transverse momentum and a 5.8 TeV/c forward momentum yielding a rather small
opening angle of $(20 \times 10^6) / (5.8 \times 10^{\,12})$ =   3.4 $\mu$\,rad.  So we dilute the radiation with a roller coaster motion\,\cite{Neuffer3} in FODO lattice  arcs 
similar to the Tevatron helical lattice motion\,\cite{Goderre}.  A rise or fall of 1\,cm over a distance of 20\,m leads to a 500 $\mu$\,rad angle, 150 times
larger than angle from muon decay.  The radiation dose falls by this factor to 4 mrem\,/year, equivalent to eating one banana a day.  Vertical bumps are used to phase shift the roller coaster motion
a few times a day. A similar phase shifting, helical lattice  is used in straight sections.

Now we see if beam power and energy losses in magnets are plausible. The same magnets are used for muon acceleration as were used for the $p \, \bar{p}$ machine.
The beam power for $4 \times 10^{\,12}$ 17.5 TeV  muons is: 

\begin{equation}
P = {{(4 \times 10^{\,12}) ( 17.5 \times 10^{\,12}) (1.6 \times 10^{-19})} \over {0.364 \, \hbox{s}}} = 31 \, \hbox{MW.}
\end{equation}

The ultra low carbon steel eddy current losses are\,\cite{Sasaki}:

\begin{equation}
P = \hbox{[Duty Factor][Volume]}{{(2\pi\,f\,B\,w)^2}\over{24\rho}}
= 14\,\hbox{MW,}
\end{equation}

where the duty factor due to the flat top is 0.30, the steel volume is 15,\,000 m$^3$, the frequency is 9 Hz, the magnetic field averages 0.9 Tesla in the steel,
the lamination width is 0.0005 m, and the resistivity of the steel is $9.6 \times 10^{-9}$ n$\Omega$-m.  Using the Steinmetz Law\,\cite{Dawes}
the hysteresis loss is:

\begin{equation}
\hbox{Energy\,/\,cycle} = (.001)(9000 \, \hbox{gauss})^{1.6}  = 2100 \, \hbox{ergs\,/\,cc}
\end{equation}

\begin{equation}
P\!=\!\hbox{(Vol\,/cycle) (2100 ergs\,/cc) (10$^{-7}$\,joules\,/erg) = 9 MW,}
\end{equation}

where the volume is 15,\,000\,m$^3$ times $10^6$ cc\,/m$^3$ and the cycle time is 0.364 seconds.
Tests of  YBCO superconductor ramping at 9 Hz are showing  progress\,\cite{Piekarz}.

Next, we accelerate muons\,\cite{Summers} in a Fermilab site filler ring to 1.75 TeV, and then to 17.5 TeV in the 233 km circumference ring
using 2 Tesla dipoles, 250 GV of superconducting RF, and 63 orbits. Phase\,/frequency locked magnetrons\,\cite{Dexter} might supply power for the RF,
if they can be developed as a more efficient alternative to klystrons.

\begin{equation}
{\rm{SURVIVAL}} = \prod_{N=1}^{63} \exp\left[{{-2\pi{R}\,m_{\mu^{\pm}}} \over {[1625 + (250\,N)]\,c\,\tau}}\right] = 71\%
\end{equation}

A final focus system has been worked out for a 30 TeV, round beam, muon collider\,\cite{Raimondi2000}.
The IP beta function, $\beta^{\,*}$, is 0.48\,cm. Quadrupole gradients are below 400 T/m and peak fields are below 15 T.
Twelve meters is kept free for a detector.  Total length of this final focus system is 2 km. Initially the acceleration ring is used as a 35 TeV collider with 
two detectors to give a luminosity of:

 \begin{equation}
 L = {{\gamma \, N^{\,2} f_0} \over {4 \pi  \epsilon^N \beta^*\rule{0pt}{9pt}}}  = {{165,000 \, (2 \times 10^{\,12})^2 \ 2575} \over {4 \pi \, (25 \times 10^{\,-4}\,{\rm{cm}}) \, 0.48}} = {{1.1 \times 10^{\,\,35}}
 {\rm{\,cm}^{\,{-2}} \, s^{-1}}}
 \end{equation}

A smaller collision ring with higher field dipole magnets, higher collision rates, and higher luminosity could be added as an upgrade.
Progress is being made on $\epsilon^N =$ 25 mm-mrad cooled muon bunches but more work remains\,\cite{Cool}.

%http://mafurman.lbl.gov/LBNL-38563.pdf

\bigskip
\bigskip
\centerline{\bf \large SUMMARY}
\medskip

\noindent
A site filler ring at  Fermilab permits 240 GeV $e^+ e^-$ and 3.5 TeV $\mu^+\mu^-$ colliders.
The 233 km tunnel would be  used by a series of  machines sequentially over many decades. The tunnel  would first be filled  with 140 gauss dipoles for the 240 GeV $e^+ e^-$ machine
aimed at producing one million  $e^+ e^- \!\to Z^{\,0} h^0$ reactions per year. 
%A crab waist crossing\,\cite{Raimondi2006} may help extend the energy reach of $e^+ e^-$ beyond LEP.
The Tevatron $\bar{p}$ source gives a competitive event rate with a 40 TeV energy frontier hadron collider based on 2\,T dipole magnets.  
Muon acceleration from 1.75 to 17.5 TeV with 250 GV of RF and 2\,T dipole magnets with 9\,Hz ramped superconducting coils looks promising.
Neutrino radiation might be rasterized. A 35 TeV muon collider would have 70x the center of mass energy of the ILC, while using half the RF of the ILC.
%One option for powering these rings is a subcritical, thorium\,/nuclear waste reactor\,\cite{Guenevere}
%driven by a fixed field, alternating gradient accelerator.

\bigskip
\centerline{\bf \large ACKNOWLEDGMENT}
\medskip

\noindent
Many thanks S.\,Berg, A.\,Blondel, J.\,Byrd, H.\,Frisch,
E. Gianfelice\,-Wendt, K.\,Gollwitzer, S.\,Hansen, D.\,Hazel\-ton, C.~Johnstone, J.\,Kraus, N.\,Mokhov, S.\,Mrenna, R.\,Palmer, M.\,Ross, P. Rubinov,
A.\,Tollestrup, and F. Zimmermann for useful discussions.
This work was  supported by NSF PHY-1068052, NSF 757938, and DOE DE-FG05-91ER40622.

\bigskip
\centerline{\bf \large REFERENCES}
\vspace{-10mm}

\renewcommand\refname{}

%\begin{thebibliography}{9}   % Use for  1-9  references

}


\begin{thebibliography}{99} % Use for 10-99 references

\bibitem{CNA}
%CNA Consulting Engineers (Hatch-Mott-MacDonald),
%``Estimate of Heavy Civil Underground Construction Costs for a Very Large Hadron Collider in Northern Illinois,"
VLHC-2001-CNA-Report, http:\,//vlhc.org/cna/cna\_report.pdf





\bibitem{Ambrosio}
G. Ambrosio {\it et al.}\,(VLHC), 
%``Design Study for a Staged Very Large Hadron Collider," 
Fermilab\,-TM-2149 (2001).

\bibitem{Lyons}
G. T. Lyons, Master's Thesis, 
%``A 233 km Circumference Tunnel for $e^+ e^-$, $p \bar{p}$, and $\mu^+ \mu^-$ Colliders,"
arXiv:1112.1105; \newline
G. T. Lyons {\it et al.,} IPAC-2012-TUPPR008.

\bibitem{Sen}
T. Sen and J. Norem, 
%``A Very Large Lepton Collider in the VLHC Tunnel,"
Phys. Rev. ST Accel. Beams {\bf 5} (2002) 031001.

\bibitem{Neuffer} D. Neuffer, AIP Conf. Proc. {\bf 156} (1987) 201; \newline
D. V. Neuffer and R. B. Palmer, Conf. Proc. C940627 (1994) 52; \newline
D. J. Summers,  Bull. Am. Phys. Soc. {\bf 39} (1994) 1818; \newline
J. Gallardo {\it et al.,} Snowmass 1996, BNL-52503; \newline
C.\,Ankenbrandt {\it et al.,} Phys.\,Rev.\,ST Accel. Beams {\bf 2} (1999) 081001; \newline
M.\,Alsharo'a {\it et al.,} Phys.\,Rev.\,ST\,Accel. Beams {\bf 6} (2003) 081001; \newline
A. Moretti {\it et al.} Conf. Proc. C0408164 (2004) 271; \newline
R.\,Palmer {\it et al.,} Phys.\,Rev.\,ST Accel. Beams {\bf 12} (2009) 031002.

\bibitem{Boson}
ATLAS Collaboration, ``Observation of a 126 GeV Boson,'' (Jul 2012); \newline
CMS Collaboration, ``Observation of a 125 GeV Boson,'' (Jul 2012).

\bibitem{Blondel}
A. Blondel and F. Zimmermann,
%``A High Luminosity $e^+ e^-$ Collider in the LHC Tunnel to Study the Higgs Boson,"
arXiv:1112.2518; \newline
A. P. Blondel {\it et al.,} IPAC-2012-TUPPR078; \newline
K. Yokoyo, ``Scaling of High-Energy $e^+ e^-$ Ring Colliders,'' KEK Accelerator Seminar, 15 Mar 2012.

\bibitem{Neuffer2}
D. Neuffer, ``A 125 GeV Muon Higgs Collider,'' Advanced Accelerator Concepts Workshop, Austin, TX, 10-15 Jun 2012; \newline
R. Raja and A. Tollestrup, Phys. Rev. {\bf D58} (1998) 013005; \newline
V. Barger, M. Berger, J. Gunion, and T. Han, Phys Rept. {\bf 286} (1997) 1.

\bibitem{Raimondi2006}
P.  Raimondi,
%``Status on Super-B Effort,"
Conf. Proc.\,C0606141 (2006) 104; \newline
P.\,Raimondi, D.\,Shatilov, and M.\,Zobov,
%``Beam-Beam Issues for Colliding Schemes with Large Piwinski Angle and Crabbed Waist,"
physics\,/0702033.

\bibitem{Brau}
J. Brau {\it et al.,}
%``International Linear Collider Reference Design Report,"
SLAC-R-857 (2007).

\bibitem{Lee}
S. Y. Lee. ``Accelerator Physics'' (2012) 425; \newline
H.\,Wiedemann, ``Particle Accelerator Physics I'' (1999) 405.

\bibitem{Bossert}
R. Bossert {\it et al.,}
%``Optimization and Test of a 120mm LARP Nb$_3$Sn Quadrupole Coil using Magnetic Mirror Structure,"
%FERMILAB-CONF-11-427-TD.
IEEE  Trans. Appl. Supercond. {\bf 22} (2012) 4003404.

\bibitem{Telnov}
V. I. Telnov, arXiv:1203.6563.

\bibitem{Machida}
S. Machida {\it et al.,} Nature Phys. {\bf 8} (2012) 243.

\bibitem{Shirkoohi}
G.\,Shirkoohi and  M.\,Arikat,
%``Anisotropic Properties of High Permeability Grain-Oriented 3.25\% Si-Fe Electrical Steel,"
IEEE Trans. Magnetics {\bf 30} (1994) 928; \newline
D. J. Summers {\it et al.,} 
%``Test of a 1.8 Tesla, 400 Hz Dipole for a Muon Synchrotron,"
IPAC-2012-THPPD020.

\bibitem{Laeger}
H. Laeger {\it et al.,}
%``Production of the Soft Magnetic Steel Laminations for the LEP Dipole Magnets,"
IEEE Trans. Magnetics {\bf 24}  (1988) 835; \newline
http:\,//www.cmispecialty.com/31558\_CMI-B\_Data\_Sheet.pdf

\bibitem{Holmes}
S. Holmes {\it et al.,}
%``Overview of the Tevatron Collider Complex: Goals, Operations and Performance,"
JINST {\bf 6} (2011) T08001.
   
 \bibitem{Pasquinelli}
 R. J. Pasquinelli {\it et al.,} 
 %``Progress in Antiproton Production at the Fermilab Tevatron Collider,"
 Fermilab-Conf-09-126-AD (2009). 

\bibitem{Stelzer}
T.  Stelzer and W.  Long,
%``Automatic Generation of Tree Level Helicity Amplitudes,"
Comp. Phys. Comm. {\bf 81} (1994) 357; \newline
J. Alwall {\it et al.,}
%``MadGraph/MadEvent v4: The New Web Generation,"
JHEP {\bf 0709} (2007) 028; \newline
J. Alwall {\it et al.,}
%``MadGraph 5:  Going Beyond,"
JHEP {\bf 1106} (2011) 128.

\bibitem{King}
B. J. King, 
%``Potential Hazards from Neutrino Radiation at Muon Colliders,"  
physics\,/\,9908017.

\bibitem{Neuffer3}
D. Neuffer, CERN-YELLOW-99-12; \newline
N. Mokhov and A. Van Ginneken,  J. Nucl. Sci. Tech. {\bf S1} (2000) 172.

\bibitem{Goderre}
 G. P. Goderre and  E. Malamud,
Conf. Proc. C8903201 (1989) 1818.

\bibitem{Sasaki}
H. Sasaki, 
%``Magnets for Fast Cycling Synchrotrons,"
KEK-PREPRINT-91-216.

\bibitem{Dawes}
C.\,L.\,Dawes,
``A Course in Electrical Engineering'' (1920) 182.

\bibitem{Piekarz}
H. Piekarz {\it et al.,} 
IEEE  Trans. Appl. Supercond. {\bf 22} (2012) 5800105; \newline
H. Piekarz {\it et al.,} IEEE Trans. Appl. Supercond. {\bf 20} (2010) 1304.

\bibitem{Summers}
D. J. Summers {\it et al.,} 
% ``Muon Acceleration to 750 GeV in the Tevatron Tunnel for a 1.5 TeV $\mu^+ \mu^-$ Collider,"
arXiv:0707.0302.

 \bibitem{Dexter}
 A. C. Dexter {\it et al.}
% ``First Demonstration and Performance of an Injection Locked Continuous Wave Magnetron to Phase Control a Superconducting Cavity,"   
 Phys. Rev. ST Accel. Beams {\bf 14} (2011) 032001; \newline
M. Neubauer {\it et al.,} IPAC-2011-MOPC140.


\bibitem{Raimondi2000}
P. Raimondi and   F. Zimmermann, 
%``Performance of a Compact Final Focus System for a 30\,-TeV Muon Collider,"
SLAC-REPRINT-2000-168; \newline
P. Raimondi and A. Seryi, Phys. Rev. Lett. {\bf 86} (2002) 3779.
 
\bibitem{Cool}
D. Neuffer, Part. Accel. {\bf  14} (1983) 75; \newline
R. Fernow and J. Gallardo, Phys. Rev. {\bf E52} (1995) 1039; \newline
K.-J. Kim and C.-x. Wang, Phys. Rev. Lett. {\bf 85}  (2000) 760; \newline
G. Penn and J. S. Wurtele, Phys. Rev. Lett. {\bf 85}  (2000) 764; \newline
R.\,Palmer {\it et al.,} Phys.\,Rev.\,ST Accel. Beams {\bf 8} (2005) 061003; \newline
R.\,Palmer {\it et al.,} arXiv:0711.4275;  \newline
R. Palmer {\it et al.,} PoS NUFACT08 (2008) 019; \newline
M.\,Bogomilov {\it et al.}\,(MICE Collaboration), JINST {\bf 7} (2012) P05009; \newline
T. Hart {\it et al.,} IPAC-2012-MOPPC046; \newline
T. Schwarz {\it et al.,} IPAC-2012-WEOBA03; \newline
Y. Shiroyanagi {\it et al.,} IPAC-2012-THPPD048.


%\bibitem{Guenevere}
%GUENEVERE, CERN Cour. {\bf 52N3} (2012) 12; \newline
%C. D. Bowman and R. P. Johnson, IPAC-2011-THOAB01; \newline
%C. Rubbia, AIP Conf. Proc. {\bf 346} (1995) 44.
  

\end{thebibliography}
\end{document}